\documentclass{article}
\usepackage[latin9]{inputenc}
\usepackage{geometry}
\usepackage{graphicx}
\usepackage{subfig}
\usepackage{amsmath}
\usepackage{amssymb}
\usepackage[unicode=true,pdfusetitle,
 bookmarks=true,bookmarksnumbered=false,bookmarksopen=false,
 breaklinks=false,pdfborder={0 0 1},backref=section,colorlinks=false]
 {hyperref}
\usepackage{breakurl}

\makeatletter

\usepackage{float}
\usepackage{grffile}
\usepackage{breakurl}

\usepackage{fullpage}\usepackage[justification=centering]{caption}\usepackage[all]{hypcap}\usepackage{microtype}

\makeatother

\begin{document}

\title{The scissors mode from a different perspective}

\author{Matthew Harper, Larry Zamick\\
 \textit{Department of Physics and Astronomy, Rutgers University,
Piscataway, New Jersey 08854}\\
 }

\date{\today}

\maketitle 

\begin{abstract}
The scissors mode, a magnetic dipole excitation-mainly orbital is
usually discussed in terms of a transition from a $J=0^{+}$ ground
state to a $J=1^{+}$ excited state. This is understandable because
it follows from the way the experiment is performed-e.g. inelastic
electron scattering. Here however, we start with the excited $1^{+}$
state and consider all possible transitions to $J=0^{+}$,1$^{+}$
and 2$^{+}$states with final isospins. There is a larger transition
to the 0$_{2}^{+}$state than to ground. This has a much richer structure.We
note that the ``sum of sums'' is independent of the interaction. 
\end{abstract}

\section{Introduction}

\indent\indent In a collective picture the scissors mode is an orbital
magnetic dipole excitation, in which the deformed proton symmetry
axis vibrates against the corresponding axis of the neutrons. Some
early discussions of this mode are contained by Richter's group, Bohle
et al. \cite{Bohle and Richter,Bohle and Kuchler}. 
In 2010 there was an extensive review of scissors
modes by K. Heyde at al, \cite{Heyde}. This has stimulated research by many different groups , both theoretical and experimental.  It will not be practical to mention all of those that are referred to in the review article but we selected some that show the variety of approaches. These are listed in references \cite{Suzuki}-\cite{LoIudice} More recently, there has been work on M1 excitations by J. Beller et al. \cite{Beller} in which the initial state has $I=1^+$. This is of great relevance to the theme of the present work. 

In all the experiments which are mainly inelastic electron scattering,
one starts with the $J=0^{+}$ ground state and considers excitations
to $J=1^{+}$ states. The supporting calculations follow suit. However,
since there are no practical constraints for theory, we will here
start with the $J=1^{+}$ scissors mode state and follow the various
branches to which it can connect. Now we can go not only from $J=1^{+}$
to $J=0^{+}$ but also $J=1^{+}$ to $J=2^{+}$ which gives a much
richer spectrum.

This work can be regarded as an extension of previous work by the
authors \cite{Harper}. In that work the main focus was on selection rules
with a $J=0$ $T=1$ pairing interaction i.e. why certain B(M1)'s
vanish. In this work we will make quantitative comparisons of the
non-vanishing strengths with different interactions. For example, there has been considerable work on $J_{max}$ pairing by Zhao and Arima \cite{Zhao}, Cederwall \cite{Cederwall}, Xu et al. \cite{Xu}, Fu et al. \cite{Fu} Zamick and Escuderos \cite{Zamick and Escuderos}, Hertz-Kintish and Zamick \cite{Hertz-Kintish}.

\section{B(M1) Results for Various Interactions}

\indent\indent We present results in Tables II through XXII, which are
$^{44}$Ti $I=1$ to 0, $^{44}$Ti $I=1$ to 2, $^{46}$Ti $I=1$
to 0, and $^{46}$Ti $I=1$ to 2. As well as, $^{44}$Ti $I=1$ to 1 , $^{46}$Ti $I=1$ to 1. There are four interactions used- $J$=0 $T$=1 Pairing, Q.Q, MBZE \cite{Escuderos} and $J_{max}$
$T$=0 pairing. These are represented by 8 numbers (7 independent), corresponding
to two nucleons coupled to $J$=0 to $J$=7. Here they are:\\

\subsection*{Table I. Matrix Elements for the Interactions}
\centering $\begin{array}{c|cccccccc}
\hline
J=0\text{ Pairing}& -2& 0& 0& 0& 0& 0& 0& 0\\
\text{Q.Q}& 0& 0.4096& 1.1471& 2.0483& 2.8677& 3.2744& 2.8677& 1.1471\\
\text{MBZE}& 0& 0.6111& 1.5863& 1.4904& 2.8153& 1.5101& 3.2420& 0.6163\\
J_{\text{max}} \text{ Pairing}& 0& 0& 0& 0& 0& 0& 0& -2\\
\hline
\end{array}$\\

\raggedright

In some cases, in order to remove degeneracies with schematic interactions
we add -1.00 MeV to all the odd $J$, $T$=0 matrix elements. If we did
not do this, then states of different isospins would be degenerate
and arbitrary mixtures of these states would appear in the computer
output. This trick pushes up states of higher isopin to higher energies, but leaves the
energies of lower isospin unchanged. We call these new energies shifted. These higher isospin states in $^{44}$Ti are indicated with a star $(*)$ for T=1 and two stars $(**)$ for T=2. Similarly, higher isospin states for $^{46}$Ti are given one star $(*)$ for T=2 and two stars $(**)$ for T=3. We give the seniority, isospin, and reduced isospin for the pairing interactions so we do not use the star notation for labeling the states.

We also present the results in various figures. 
All B(M1)'s are in units of $(\mu_N)^2$.

\subsection*{Table II. Pairing B(M1) $^{44}$Ti $I$=1 to $I$=0}

\centering $\begin{array}{cc|ccc|cc}
\hline \text{State}(v,T,t) & I=1 & 210 & 411 & 411\\
I=0 & \text{Unshifted Energy} & 1.500 & 2.250 & 2.250 & \text{sum}\\
\hline 000 & 0.000 & 2.6996 & 0 & 0 & 2.6996\\
020 & 0.750 & 8.0995 & 0 & 0 & 8.0995\\
400 & 2.250 & 1.9300 & 0.1117 & 2.8922 & 4.9339\\
400 & 2.250 & 0.8986 & 7.7693 & 1.9187 & 10.4966\\
\hline  & \text{sum} & 13.6277 & 7.7910 & 4.8109 & 26.2296
\end{array}$\\
 \raggedright

\subsection*{Table III. Pairing B(M1) $^{44}$Ti $I$=1 to $I$=2}

\centering $\begin{array}{cc|ccc|cc}
\hline \text{State}(v,T,t) & I=1 & 210 & 411 & 411\\
I=2 & \text{Unshifted Energy} & 1.500 & 2.250 & 2.250 & \text{sum}\\
\hline 201 & 1.000 & 2.6015 & 0.4505 & 1.3857 & 4.4377\\
400 & 1.250 & 44.8541 & 3.7086 & 3.5942 & 52.1569\\
400 & 1.750 & 1.6209 & 6.3448 & 1.8866 & 9.8523\\
400 & 2.250 & 6.0571 & 4.7158 & 9.1692 & 19.9421\\
221 & 2.250 & 0 & 0 & 0 & 0\\
411 & 2.250 & 0 & 0 & 0 & 0\\
411 & 2.250 & 0 & 0 & 0 & 0\\
411 & 2.250 & 13.0086 & 2.2518 & 6.9270 & 22.1874\\
422 & 2.250 & 0.00005 & 21.4800 & 1.0920 & 22.5721\\
\hline  & \text{sum} & 68.1423 & 38.9515 & 24.0547 & 131.1483
\end{array}$\\
 \raggedright

\subsection*{Table IV. Pairing B(M1) $^{46}$Ti $I$=1 to $I$=0}

\centering $\begin{array}{cc|ccccccc|cc}
\hline \text{State}(v,T,t) & I=1 & 220 & 411 & 411 & 421 & 421 & 611 & 611\\
I=0 & \text{Unshifted Energy} & 1.7500 & 2.0000 & 2.0000 & 2.5000 & 2.5000 & 2.7500 & 2.7500 & \text{sum}\\
\hline 010 & 0 & 1.0799 & 0 & 0 & 0 & 0 & 0 & 0 & 1.7099\\
030 & 1.2500 & 9.7200 & 0 & 0 & 0 & 0 & 0 & 0 & 9.7200\\
410 & 2.2500 & 2.4344 & 2.8794 & 0.0491 & 0.5611 & 0.4150 & 0 & 0 & 6.3390\\
410 & 2.7500 & 0.3947 & 0.7573 & 5.7648 & 0.1157 & 2.0588 & 0 & 0 & 6.3390\\
611 & 2.7500 & 0 & 1.0423 & 0.0987 & 3.1539 & 0.2640 & 2.3989 & 0.6317 & 9.0913\\
611 & 2.7500 & 0 & 0.0049 & 0.1721 & 0.0858 & 0.4450 & 0.0001 & 1.7267 & 7.5895\\
\hline  & \text{sum} & 13.6290 & 4.6839 & 6.0847 & 3.9165 & 3.1828 & 2.3990 & 2.3584 & 36.2543
\end{array}$\\

\raggedright

\subsection*{Table V. Pairing B(M1) $^{46}$Ti $I$=1 to $I$=2}

\centering $\begin{array}{cc|ccccccc|c}
\hline \text{State}(v,T,t) & I=1 & 220 & 411 & 411 & 421 & 421 & 611 & 611\\
I=2 & \text{Unshifted Energy} & 1.7500 & 2.0000 & 2.0000 & 2.500 & 2.500 & 2.7500 & 2.7500 & \text{sum}\\
\hline 211 & 1.0000 & 1.3712 & 0.9874 & 0.3326 & 0.0005 & 0.0019 & 0 & 0 & 2.6936\\
211 & 1.0000 & 0.1715 & 0.4367 & 0.1472 & 0.0813 & 0.3238 & 0 & 0 & 1.1605\\
221 & 1.5000 & 2.5716 & 2.2323 & 0.7524 & 0.0222 & 0.0883 & 0 & 0 & 5.6668\\
412 & 1.5000 & 0 & 0.0916 & 1.5360 & 0.0607 & 0.4819 & 0 & 0 & 2.1702\\
411 & 2.0000 & 0 & 0.0847 & 0.0914 & 0.5024 & 0.0261 & 0.4364 & 0.0065 & 1.1475\\
411 & 2.0000 & 0 & 0.0041 & 0.0186 & 0.0014 & 0.0668 & 1.5191 & 0.0152 & 1.6244\\
422 & 2.0000 & 0 & 0.2746 & 4.6069 & 0.1821 & 1.4454 & 0 & 0 & 6.5090\\
410 & 2.2500 & 12.1303 & 0.0646 & 1.6850 & 0.0832 & 0.5004 & 0 & 0 & 14.4635\\
410 & 2.2500 & 2.9785 & 3.5617 & 0.1189 & 0.6431 & 0.5838 & 0 & 0 & 7.8860\\
410 & 2.2500 & 5.3986 & 0.4668 & 2.4445 & 0.0273 & 0.9432 & 0 & 0 & 9.2804\\
231 & 2.2500 & 2.0572 & 0 & 0 & 1.1354 & 4.5230 & 0 & 0 & 7.7156\\
421 & 2.5000 & 0 & 0.1804 & 0.0338 & 0.6237 & 0.0188 & 0.6123 & 0.0630 & 1.5320\\
421 & 2.5000 & 0 & 0.0862 & 0.2962 & 0.8883 & 0.2597 & 5.2534 & 0.0019 & 6.7857\\
611 & 2.7500 & 0 & 2.1377 & 0.2523 & 6.7325 & 0.4370 & 2.3618 & 0.0555 & 11.9768\\
611 & 2.7500 & 0 & 0.2654 & 0.0135 & 0.4044 & 0.4321 & 0.1597 & 0.8390 & 2.1141\\
611 & 2.7500 & 0 & 0.0367 & 0.1344 & 0.0050 & 0.5082 & 7.1099 & 1.4178 & 9.2120\\
611 & 2.7500 & 0 & 0.0375 & 0.0024 & 0.1070 & 0.0127 & 0.0873 & 0.0461 & 0.2930\\
611 & 2.7500 & 0 & 0.1215 & 1.3291 & 0.3483 & 4.0036 & 0.00007 & 5.7321 & 11.5347\\
\hline  & \text{sum} & 26.6789 & 11.0699 & 11.8488 & 14.6567 & 13.7952 & 17.5400 & 8.1771 & 103.7666
\end{array}$\\
 \raggedright

\subsection*{Table VI. Q.Q B(M1) $^{44}$Ti $I$=1 to $I$=0}

\centering $\begin{array}{cc|ccc|cc}
\hline  & I=1 & 1_{1} & 1_{2} & 1_{3}\\
I=0 & \text{Unshifted Energy} & 3.3648^* & 6.3405^* & 9.5620^* & \text{sum}\\
\hline 0_{1} & 0.0000 & 1.3174 & 0.0015 & 0.0007 & 1.3196\\
0_{2} & 3.6031^{**} & 1.8021 & 6.1454 & 0.1535 & 8.1010\\
0_{3} & 7.5748 & 0.1833 & 9.0414 & 0.9530 & 10.1777\\
0_{4} & 10.9236 & 0.0414 & 0.0577 & 6.5332 & 6.6323\\
\hline  & \text{sum} & 3.3442 & 15.2460 & 7.6404 & 26.2306
\end{array}$\\
 \raggedright

\subsection*{Table VII. Q.Q B(M1) $^{44}$Ti $I$=1 to $I$=2}
\centering 
$\begin{array}{cc|ccc|ccc}
\hline  & I=1 & 1_{1} & 1_{2} & 1_{3}\\
I=2 & \text{Unshifted Energy} & 3.3648^* & 6.3405^* & 9.5620^* & \text{sum}\\
\hline 2_{1} & 0.9655 & 2.4898 & 0.0111 & 0.0016 & 2.5025\\
2_{2} & 3.6015^* & 0 & 0 & 0 & 0\\
2_{5} & 4.7502^* & 0.1735 & 20.6912 & 1.3251 & 22.1898\\
2_{3} & 6.4691 & 13.7051 & 8.2795 & 1.2061 & 23.1907\\
2_{6} & 7.5695^* & 0 & 0 & 0 & 0\\
2_{8} & 7.6179^{**} & 0.1271 & 0.8452 & 21.6001 & 22.5724\\
2_{4} & 7.7501 & 0.0545 & 46.3395 & 0.6632 & 47.0572\\
2_{9} & 9.7351^* & 0 & 0 & 0 & 0\\
2_{7} & 10.4893 & 0.1723 & 0.0767 & 13.4096 & 13.6586\\
\hline  & \text{sum} & 16.7223 & 76.2432 & 38.2057 & 131.1711
\end{array}$\\
 \raggedright

\subsection*{Table VIII. Q.Q B(M1) $^{46}$Ti $I$=1 to $I$=0}

$\begin{array}{cc|ccccccc|cc}
\hline  & I=1 & 1_{1} & 1_{2} & 1_{3} & 1_{4} & 1_{5} & 1_{6} & 1_{7}\\
I=0 & \text{Unshifted Energy} 
                 & 4.3546     & 8.1095 &8.7081^{*}&10.4611^{*}& 10.5846&10.8481 &11.6407^{*}& \text{sum}\\

\hline 
0_{1} & 0.0000      & 1.3901  & 0.0006 &0.0038    & 0         & 0      & 0      &  0.0002 & 1.3947\\
0_{2} & 6.4642      & 2.6505  & 0.0897 &2.2242    &0.0003    & 0.3008 & 0.0015 &   0.0161 & 5.2831\\
0_{3} & 7.9741      & 0.1986  & 5.0379 &0.3734    &  0.1137    & 0.0310 & 0.2988 & 0.00004 & 6.0534\\
0_{6} & 9.7237^{**} & 0       & 0      &6.8081    &0         & 0      & 0      &   2.9145 & 9.7226\\
0_{4} & 10.7392     & 0.0191  & 0.4441 &0.2985    &2.9265    & 0.1054 & 4.0417 &   0.8731 & 8.7084\\
0_{5} & 12.5438     & 0.000099& 0.0097 &0.0013    &0.0020    & 0.8983 & 0.0079 &   4.1814 & 5.1007\\
\hline  & \text{sum}& 4.2584  & 5.5820 &9.7093    &3.0425    & 1.3355 & 4.3499 &   7.9853 & 36.2629
\end{array}$ \raggedright

\subsection*{Table XI. Q.Q B(M1) $^{46}$Ti $I$=1 to $I$=2}

\centering $\begin{array}{cc|ccccccc|c}
\hline  & I=1 & 1_{1} & 1_{2} & 1_{3} & 1_{4} & 1_{5} & 1_{6} & 1_{7}\\
I=2 & \text{Unshifted Energy} & 
                     4.3546 & 8.1095 &8.7081^*&10.4611^*& 10.5846&10.8481 &11.6407^*& \text{sum}\\
\hline 
2_{1} & 0.8630     & 0.6034 & 0.0445 &0.0011 & 0.0030 &0      &0.0001  &    0.0002  & 0.6523\\
2_{2} & 3.5162     & 1.4891 & 0.3974 &0.0151 & 0.0182 &0.0005 &0.0050  &    0.0017  & 1.9270\\
2_{3} & 4.2764     & 1.8998 & 0.3245 &0.1051 & 0.0023 &0.0045 &0.0495  &    0.0017  & 2.3874\\
2_{4} & 6.2720     & 0.1660 & 0.0521 &0.1476 & 0.0395 &0.0276 &0.1994  &    0.0026  & 0.6348\\
2_{5} & 7.2633     & 0.0486 & 0.5793 &0.2012 & 0.3204 &0.0569 &0.0156  &    0.0226  & 1.2446\\
2_{14}& 7.3607^*   & 2.3182 & 0.0359 &1.9267 & 0.0346 &0.4418 &0.0221  &    0.0027  & 4.7820\\
2_{6} & 7.7478     & 1.1676 & 0.5558 &10.2928& 0.0683 &0.2179 &0.0103  &    0.0004  & 12.3131\\
2_{7} & 8.5830     & 0.0120 & 5.2029 &0.0168 & 2.4499 &0.2874 &0.0202  &    0       & 7.9892\\
2_{15}& 9.6011^*   & 0.0372 & 3.2940 &0.5506 & 0.5938 &0.3825 &3.3463  &    0.0549  & 8.2593\\
2_{8} & 9.6672     & 0.0428 & 0.0018 &5.5599 & 0.2733 &2.7290 &0.1841  &    2.1744  & 10.9653\\
2_{16}& 9.8751^*   & 0.1340 & 0.0460 &0.0143 & 0.8130 &4.0584 &0.0469  &    0.0597  & 5.1723\\
2_{9} & 10.5511    & 0.0154 & 0.0132 &0.0092 & 4.2207 &3.1012 &0.1246  &    0.1755  & 7.6598\\
2_{18}&10.8708^{**}&      0 &      0 &0.2478 & 4.0121 &0      &0       &    3.4565  & 7.7164\\
2_{10}& 11.2619    & 0.0079 & 0.0528 &1.0658 & 0.0045 &3.7622 &0.0004  &    4.0527  & 8.9463\\
2_{11}& 11.3626    &      0 & 0.0340 &0.0075 & 0.1386 &0.3283 &10.7713 &    0.1488  & 11.4285\\
2_{17}& 12.1399^*  & 0.0005 & 0.00004&0.0345 & 0.5456 &0.1533 &0.0738  &    1.4694  & 2.27714\\
2_{12}& 12.4314    & 0.0004 & 0.0009 &0.0935 & 0.4677 &1.4645 &0.0974  &    6.3983  & 8.5227\\
2_{13}& 12.8660    &      0 & 0.0084 &0.0033 & 0.1017 &0.0005 &0.0112  &    0.7700  & 0.8951\\
\hline &\text{sum} & 7.9429 & 10.6435&20.2928& 14.1072&17.0165&14.9782 &    18.7921 & 103.7732
\end{array}$

\raggedright

\subsection*{Table X. MBZE B(M1) $^{44}$Ti $I$=1 to $I$=0}

\centering

$\begin{array}{cc|ccc|cc}
\hline  & I=1 & 1_{1} & 1_{2} & 1_{3}\\
I=0 & \text{Unshifted Energy} & 5.66864^{*} & 7.58685^{*} & 9.72619^{*} & \text{sum}\\
\hline 0_{1} & 0.00000 & 1.18248 & 0.17056 & 0 & 1.35304\\
0_{2} & 5.58610 & 0.13111 & 5.29543 & 0.05642 & 5.48296\\
0_{3} & 8.28402^{**} & 1.95508 & 6.07014 & 0.07579 & 8.10101\\
0_{4} & 8.7875 & 0.17022 & 1.73958 & 9.38455 & 11.29435\\
\hline  & \text{sum} & 3.43889 & 13.27571 & 9.51676 & 26.2314
\end{array}$

\raggedright

\subsection*{Table XI. MBZE B(M1) $^{44}$Ti $I$=1 to $I$=2}

\centering $\begin{array}{cc|ccc|cc}
\hline  & I=1 & 1_{1} & 1_{2} & 1_{3}\\
I=2 & \text{Unshifted Energy} & 5.66864^{*} & 7.58685^{*} & 9.72619^{*} & \text{sum}\\
\hline 2_{1} & 1.16313 & 1.34744 & 0.42560 & 0.04716 & 1.82020\\
2_{2} & 4.95650 & 12.97910 & 1.30523 & 0.27252 & 14.55685\\
2_{3} & 5.23665^{*} & 0 & 0 & 0 & 0\\
2_{4} & 7.81197 & 0 & 0 & 0 & 0\\
2_{5} & 7.82336 & 1.09707 & 37.57634 & 12.68482 & 51.35823\\
2_{6} & 7.96963 & 1.53883 & 10.26440 & 6.83864 & 18.64187\\
2_{7} & 9.26771^{*} & 0 & 0 & 0 & 0\\
2_{8} & 9.87032^{**} & 0.09840 & 16.68741 & 5.40033 & 22.18614\\
2_{9} & 11.88190^{**} & 0.11349 & 0.11945 & 22.34037 & 22.57331\\
\hline  & \text{sum} & 17.19433 & 66.37843 & 47.58384 & 131.15657
\end{array}$\\
 \raggedright

\subsection*{Table XII. MBZE B(M1) $^{46}$Ti $I$=1 to $I$=0}

\centering $\begin{array}{cc|ccccccc|cc}
\hline  & I=1 & 1_{1} & 1_{2} & 1_{3} & 1_{4} & 1_{5} & 1_{6} & 1_{7}\\
I=0 & \text{Unshifted Energy} & 3.65521 & 6.05887 & 7.78516 & 8.73868 & 9.46213^{*} & 10.61597^{*} & 11.36444^{*} & \text{sum}\\
\hline 0_{1} & 0.00000 & 0.55962 & 0.01755 & 0.00592 & 0.07044 & 0.13076 & 0.01481 & 0.00998 & 0.80908\\
0_{2} & 4.62474 & 2.47374 & 0.18000 & 0.29729 & 0.51637 & 0.99056 & 0.14637 & 0.06077 & 4.6651\\
0_{3} & 6.27338 & 0.67490 & 4.31054 & 0.11501 & 0.15137 & 0.78449 & 0.00119 & 0.07790 & 6.1154\\
0_{4} & 7.89321 & 0.12817 & 0.46911 & 1.37366 & 0.36023 & 0.58752 & 2.16170 & 0.08942 & 5.16981\\
0_{5} & 9.31823 & 0.00351 & 0.18194 & 3.21205 & 0.38400 & 0.04369 & 0.32570 & 5.58837 & 9.73926\\
0_{6} & 13.20357^{**} & 0 & 0 & 0 & 0 & 4.6687 & 1.79917 & 3.25315 & 9.76218\\
\hline  & \text{sum} & 3.83994 & 5.15914 & 5.00393 & 1.52357 & 7.20572 & 4.44894 & 9.07959 & 36.26083
\end{array}$\\
 \raggedright

\subsection*{Table XIII. MBZE B(M1) $^{46}$Ti $I$=1 to $I$=2}

\centering $\begin{array}{cc|ccccccc|c}
\hline  & I=1 & 1_{1} & 1_{2} & 1_{3} & 1_{4} & 1_{5} & 1_{6} & 1_{7}\\
I=2 & \text{Unshifted Energy} & 3.65521 & 6.05887 & 7.78516 & 8.73868 & 9.46213^{*} & 10.61597^{*} & 11.36444^{*} & \text{sum}\\
\hline 2_{1} & 1.14826 & 0.20333 & 0.03800 & 0.00273 & 0.00135 & 0.00364 & 0.00619 & 0.02923 & 0.28447\\
2_{2} & 2.49693 & 1.20214 & 0.46560 & 0.06665 & 0.00391 & 0.08832 & 0.13831 & 0.00351 & 1.96844\\
2_{3} & 3.42179 & 1.73449 & 0.01221 & 0.22106 & 0.00967 & 0.00787 & 0.00670 & 0.03729 & 2.02929\\
2_{4} & 4.88264^{*} & 0.24575 & 1.20038 & 0.26325 & 0.00453 & 0.28514 & 0.00118 & 0.00090 & 2.00566\\
2_{5} & 5.15177 & 0.50943 & 0.76123 & 0.03848 & 0.24204 & 0.48245 & 0.17814 & 0.01875 & 2.23052\\
2_{6} & 6.15814 & 0.64804 & 0.07549 & 0.36327 & 0.03414 & 0.02360 & 1.45758 & 0.11907 & 2.72119\\
2_{7} & 6.79141 & 0.06966 & 3.30106 & 0.23216 & 0.05538 & 0.54470 & 1.07923 & 1.26442 & 6.54661\\
2_{8} & 7.25799 & 0.46379 & 1.32288 & 0.00002 & 0.23771 & 8.01263 & 0.25422 & 0.30383 & 10.59508\\
2_{9} & 7.53733 & 0.00342 & 0.03632 & 4.01174 & 0.55212 & 0.14699 & 0.45188 & 0.21191 & 5.41438\\
2_{10} & 8.22517 & 0.19893 & 0.00043 & 0.08763 & 5.40760 & 4.16620 & 0.02142 & 1.45222 & 11.33443\\
2_{11} & 8.25484^{*} & 1.62241 & 0.47339 & 0.53906 & 0.06593 & 2.14453 & 0.18933 & 0.24099 & 5.27564\\
2_{12} & 8.49974 & 0.04708 & 0.19257 & 6.07608 & 1.84693 & 0.62106 & 1.35405 & 0.32188 & 10.45965\\
2_{13} & 9.50002^{*} & 0.39393 & 1.53351 & 3.33583 & 1.81402 & 0.00121 & 0.00053 & 0.46185 & 7.54088\\
2_{14} & 9.91064 & 0.01248 & 0.00940 & 0.07097 & 2.30873 & 0.02298 & 3.36939 & 0.03948 & 5.83343\\
2_{15} & 10.18382 & 0.00954 & 0.06965 & 0.27111 & 0.94621 & 0.15988 & 1.72104 & 10.96380 & 14.14123\\
2_{16} & 10.40254^{*} & 0.05283 & 0.56227 & 0.41358 & 2.90402 & 0.01365 & 2.05851 & 0.27184 & 6.27670\\
2_{17} & 11.89813^{*} & 0.00532 & 0.00054 & 0.06234 & 0.61446 & 0.07426 & 0.44744 & 0.19571 & 1.40007\\
2_{18} & 14.78987^{**} & 0 & 0 & 0 & 0 & 0.07617 & 0.00399 & 7.63472 & 7.71488\\
\hline  & \text{sum} & 7.42257 & 10.0549 & 16.05596 & 17.04875 & 16.87528 & 12.73913 & 23.57140 & 103.76801
\end{array}$\\

\raggedright

\subsection*{Table XIV. $J_{\text{max}}$ B(M1) $^{44}$Ti $I$=1 to $I$=0}

\centering

$\begin{array}{cc|ccc|cc}
\hline  & I=1 & 1_{1} & 1_{2} & 1_{3}\\
I=0 & \text{Unshifted Energy} & 3.0851^* & 5.0769^* & 5.0769^* & \text{sum}\\
\hline 0_{1} & 1.0758 & 1.3441 & 0 & 0 & 1.3441\\
0_{2} & 5.0769 & 0.2309 & 4.6967 & 1.3398 & 6.2674\\
0_{3} & 5.0769 & 0.2687 & 5.8869 & 4.3617 & 10.5173\\
0_{4} & 5.0769^{**} & 1.1300 & 5.6054 & 4.3646 & 11.1000\\
\hline  & \text{sum} & 2.9737 & 16.189 & 7.0661 & 26.2288
\end{array}$\\

\raggedright

\subsection*{Table XV. $J_{\text{max}}$ B(M1) $^{44}$Ti $I$=1 to $I$=2}

\centering $\begin{array}{cc|ccc|cc}
\hline  & I=1 & 1_{1} & 1_{2} & 1_{3}\\
I=2 & \text{Unshifted Energy} & 3.0851^* & 5.0769^* & 5.0769^* & \text{sum}\\
\hline 
2_{1} & 1.0776       & 2.8055 & 0 & .0010 & 2.8065\\
2_{2} & 3.0518       & 10.7765 & 0.0698 & 4.6408 & 15.4871\\
2_{3} & 3.0676^*     & 0 & 0 & 0 & 0\\
2_{4} & 5.0769       & 0.4151 & 54.3381 & 2.0636 & 56.8168\\
2_{5} & 5.0769       & 0.0086 & 0.2783 & 10.9842 & 11.2711\\
2_{6} & 5.0769^*     & 0 & 0 & 0 & 0\\
2_{7} & 5.0769^*     & 0 & 0 & 0 & 0\\
2_{8} & 5.0769^{**}  & 0.68461 & 23.0100 & 8.3647 & 32.0593\\
2_{9} & 5.0769^{**}  & 0.1578 & 3.2658 & 9.2783 & 12.7019\\
\hline  & \text{sum} & 14.8682 & 80.9620 & 35.3326 & 131.1628
\end{array}$\\

\raggedright

\subsection*{Table XVI. $J_{\text{max}}$ B(M1) $^{46}$Ti $I$=1 to $I$=0}

\centering $\begin{array}{cc|ccccccc|cc}
\hline  & I=1 & 1_{1} & 1_{2} & 1_{3} & 1_{4} & 1_{5} & 1_{6} & 1_{7}\\
I=0 & \text{Unshifted Energy} & 2.4966 & 3.0668 & 4.8057 & 5.4724 & 5.1080^* & 5.6332^* & 7.0280^* & \text{sum}\\
\hline 
0_{1} & 1.0143 & 1.6533 & 0.0134 & 0.00005 & 0 & 0 & 0.0002 & 0 & 1.6670\\
0_{2} & 2.4037 & 0.0905 & 2.8076 & 0.3393 & 0.0091 & 0.0183 & 0.0161 & 0.0002 & 2.2811\\
0_{3} & 4.0284 & 1.8661 & 1.0119 & 0.0091 & 0.4326 & 0.0680 & 2.0710 & 0.0018 & 5.4605\\
0_{4} & 4.9091 & 0.0136 & 0.5472 & 4.7113 & 0.2183 & 2.8754 & 0.0207 & 0.3620 & 8.7485\\
0_{5} & 7.0280 & 0 & 0.0002 & 0.3270 & 1.4752 & 0.0037 & 0.1439 & 5.4312 & 7.3812\\
0_{6} & 7.0280^{**} & 0 & 0 & 0 & 0 & 0.0956 & 3.7398 & 5.8845 & 9.7199\\
\hline  & 
\text{sum} & 3.6236 & 4.3803 & 5.3863 & 2.1352 & 3.0610 & 5.9916 & 11.6797 & 36.2577
\end{array}$\\

\raggedright

\subsection*{Table XVII. $J_{\text{max}}$ B(M1) $^{46}$Ti $I$=1 to $I$=2}

\centering $\begin{array}{cc|ccccccc|c}
\hline  & I=1 & 1_{1} & 1_{2} & 1_{3} & 1_{4} & 1_{5} & 1_{6} & 1_{7}\\
I=2 & \text{Unshifted Energy} & 
2.4966 & 3.0668 & 4.8057 &5.1080^*& 5.4724 &5.6332^*&7.0280^*& \text{sum}\\
                 \hline 
2_{1}  & 1.0281  & 1.0052 & 0.0013 & 0.0004 &0.0008 & 0.00008&  0.00007& 0.00001& 1.0079\\
2_{2}  & 1.7145  & 1.4089 & 1.8389 &      0 &0.0055 & 0.0002 &  0.0036 & 0.00008& 3.2572\\
2_{3}  & 2.4212  & 0.0064 & 0.2088 & 0.0189 &0.0048 & 0.0114 &  0.0079 & 0.0011 & 0.2593\\
2_{4}  & 2.7178  & 0.0639 & 1.4115 & 0.0500 &0.0011 & 0.0679 &  0.0153 & 0.0019 & 1.6116\\
2_{5}  & 3.1507  & 0.0964 & 1.5097 & 0.8884 &0.0500 & 0.0149 &  0.1374 & 0.0035 & 2.7003\\
2_{6}  & 3.7368  & 0.0392 & 2.2568 & 0.0312 &4.6084 & 0.1493 &  0.0380 & 0.0006 & 7.1235\\
2_{7}  & 3.9423  & 0.1232 & 0.0827 & 0.0085 &0.7506 & 0.6391 &  0.1956 & 0.0019 & 1.8016\\
2_{12} & 4.0692^*& 2.6849 & 0.0001 & 0.0410 &0.0132 & 2.1248 &  0.7631 & 0.0010 & 5.6281\\
2_{8}  & 4.1408  & 1.2421 & 0.0179 & 0.1064 &0.0300 & 0.5087 &  4.4675 & 0.0108 & 6.3834\\
2_{9}  & 4.6429 & 0.00008 & 0.1848 & 8.8388 &0.8913 & 0.0559 &  0.0605 & 0.0219 & 10.0533\\
2_{13} & 4.8658^*& 0.0005 & 0.9376 & 4.6612 &2.0247 & 0.0229 &  0.0774 & 0.1122 & 7.8365\\
2_{10} & 5.2300 &   0.0003 & 0.4780 & 0.1015 &0.4972 & 7.5926 &  0.4851 & 0.0425 & 9.1972\\
2_{16} & 5.4124^*& 0.0017 & 0.2989 & 0.1136 &0.1323 & 3.1652 &  1.1913 & 0.5122 & 5.4152\\
2_{11} & 5.5500  & 0.0010 & 0.1784 & 0.8460 &0.4731 & 3.0674 &  7.7698 & 3.0295 & 15.3652\\
2_{14} & 7.0280       & 0 & 0.00005& 0.0058 &0.4008 & 0.1698 &  0.0383 & 1.3172 & 1.9320\\
2_{15} & 7.0280 & 0.00001 & 0.0010 & 0.3228 &0.1072 & 0.5371 &  0.0262 & 13.8722& 14.8665\\
2_{17} & 7.0280^*     & 0 & 0.0007 & 0.3348 &0.3573 & 0.0048 &  0.0049 & 0.9106 & 1.6131\\
2_{18} & 7.0280^{**}  & 0 &      0 &      0 &1.5404 &      0 &  0.6822 & 5.4929 & 7.7155\\
\hline 
 & \text{sum} & 6.6738 & 9.4072 & 16.3693 &11.8887 & 18.1321 &  15.9642 & 25.3321 & 103.7670
\end{array}$\\

\raggedright

\subsection*{Table XVIII. ALL INTERACTIONS B(M1) $^{44}$Ti $I$=1 to $I$=1}

\centering $\begin{array}{cc|ccc|c}
\hline  & I=1 & 1_{1} & 1_{2} & 1_{3}\\
I=1 & \text{Unshifted Energy} & - & - & - & \text{sum}\\
\hline 1_{1} & - & 0.1466 & 0 & 0 & 0.1466\\
1_{2} & - & 0 & 0.1466 & 0 & 0.1466\\
1_{3} & - & 0 & 0 & 0.1466 & 0.1466\\
\hline  & \text{sum} & 0.1466 & 0.1466 & 0.1466 & 0.4398
\end{array}$\\
 \raggedright

\subsection*{Table XIX. Pairing B(M1) $^{46}$Ti I=1 to I=1}

\centering $\begin{array}{cc|ccccccc|c}
\hline  & I=1 & 1_{1} & 1_{2} & 1_{3} & 1_{4} & 1_{5} & 1_{6} & 1_{7}\\
I=1 & \text{Unshifted Energy} & 1.7500 & 2.0000 & 2.0000 & 2.5000 & 2.5000 & 2.7500 & 2.7500 & \text{sum}\\
\hline 1_{5} & 1.7500 & 0.1466 & 0 & 0 & 0 & 0 & 0 & 0 & 0.1466\\
1_{1} & 2.0000 & 0 & 0.6726 & 0.0066 & 0.5047 & 0.0883 & 1.4888 & 0.00001 & 2.76104\\
1_{2} & 2.0000 & 0 & 0.0066 & 0.0328 & 0.1744 & 0.7997 & 0.3788 & 0.5271 & 1.9194\\
1_{6} & 2.5000 & 0 & 0.5047 & 0.1744 & 1.0309 & 1.6374 & 1.3985 & 0.4689 & 5.2148\\
1_{7} & 2.5000 & 0 & 0.0883 & 0.7997 & 1.6374 & 0.3968 & 4.2042 & 1.1124 & 8.2388\\
1_{3} & 2.7500 & 0 & 1.4888 & 0.3788 & 1.3985 & 4.2042 & 6.5643 & 1.0992 & 15.1338\\
1_{4} & 2.7500 & 0 & 0.00001 & 0.52713 & 0.4689 & 1.1124 & 1.0992 & 0.2795 & 3.48714\\
\hline  & \text{sum} & 0.1466 & 2.76104 & 1.9194 & 5.2148 & 8.2388 & 15.1338 & 3.48714 & 36.9016
\end{array}$\\
 \raggedright

\subsection*{Table XX. Q.Q B(M1) $^{46}$Ti I=1 to I=1}

\centering $\begin{array}{cc|ccccccc|c}
\hline  & I=1 & 1_{1} & 1_{2} & 1_{3} & 1_{4} & 1_{5} & 1_{6} & 1_{7}\\
I=1 & \text{Unshifted Energy} &  4.3546 & 8.1095 &8.7081^*&10.4611^*& 10.5846&10.8481 &11.6407^*& \text{sum}\\
\hline 1_{1} & 4.3546 & 0.1475 & 0.00001 & 0.0003 & 0.0016 & 0.0023 & 0.0012 & 0.0006 & 0.15351\\
1_{2} & 8.1095 & 0.00001 & 0.1417 & 0.0026 & 0.0084 & 0.0320 & 0.0075 & 0.0060 & 0.1982\\
1_{5} & 8.7081^* & 0.0003 & 0.0026 & 0.4191 & 0.3493 & 0.6084 & 0.2609 & 0.1632 & 1.8038\\
1_{6} & 10.4611^* & 0.0016 & 0.0084 & 0.3493 & 0.7721 & 4.4337 & 1.3977 & 0.8162 & 7.7790\\
1_{3} & 10.5846 & 0.0023 & 0.0320 & 0.6084 & 4.4337 & 6.9384 & 3.3583 & 1.4211 & 16.7942\\
1_{4} & 10.8481 & 0.0012 & 0.0075 & 0.2609 & 1.3977 & 3.3583 & 0.5204 & 0.6096 & 6.1556\\
1_{7} & 11.6407^* & 0.0006 & 0.0060 & 0.1632 & 0.8162 & 1.4211 & 0.6096 & 1.0009 & 4.0176\\
\hline  & \text{sum} & 0.15351 & 0.1982 & 1.8038 & 7.7790 & 16.7942 & 6.1556 & 4.0176 & 36.9019
\end{array}$\\
 \raggedright

\subsection*{Table XXI. MBZE B(M1) $^{46}$Ti I=1 to I=1}

\centering $\begin{array}{cc|ccccccc|c}
\hline  & I=1 & 1_{1} & 1_{2} & 1_{3} & 1_{4} & 1_{5} & 1_{6} & 1_{7}\\
I=1 & \text{Unshifted Energy} & 3.65521 & 6.05887 & 7.78516 & 8.73868 & 9.46213^* & 10.61597^* & 11.36444^* & \text{sum}\\
\hline 1_{1} & 3.65521 & 0.20475 & 0.00235 & 0.03278 & 0.17333 & 0.04688 & 0.00029 & 0.06086 & 0.52124\\
1_{2} & 6.05887 & 0.00235 & 0.12924 & 0.02634 & 0.19778 & 0.01634 & 0.06996 & 0.00190 & 0.44391\\
1_{3} & 7.78516 & 0.03278 & 0.02634 & 0.35892 & 3.47085 & 0.20043 & 1.73496 & 0.19646 & 6.02074\\
1_{4} & 8.73868 & 0.17333 & 0.19778 & 3.47085 & 6.05178 & 1.84761 & 0.33556 & 4.24016 & 16.31707\\
1_{5} & 9.46213^* & 0.04688 & 0.01634 & 0.20043 & 1.84761 & 1.09868 & 0.02403 & 0.46457 & 3.69854\\
1_{6} & 10.61597^* & 0.00029 & 0.06996 & 1.73496 & 0.33556 & 0.02403 & 0.52859 & 1.02258 & 3.71597\\
1_{7} & 11.36444^* & 0.06086 & 0.00190 & 0.19646 & 4.24016 & 0.46457 & 1.02258 & 0.20025 & 6.18678\\
\hline  & \text{sum} & 0.52124 & 0.44391 & 6.02074 & 16.31707 & 3.69854 & 3.71597 & 6.18678 & 36.90425
\end{array}$\\

\raggedright

\subsection*{Table XXII. $J_{\text{max}}$ B(M1) $^{46}$Ti $I$=1 to $I$=1}

\centering $\begin{array}{cc|ccccccc|c}
\hline  & I=1 & 1_{1} & 1_{2} & 1_{3} & 1_{4} & 1_{5} & 1_{6} & 1_{7}\\
I=1 & \text{Unshifted Energy} &
                  2.4966 & 3.0668 & 4.8057 &5.1080^* & 5.4724 &  5.6332^* & 7.0280^* & \text{sum}\\
                  \hline
                  1_{1} & 2.4966   & 0.1456 & 0.0004 & 0.0022 &0.0022 & 0.0003 &  0.0001 & 0      & 0.1508\\
                  1_{2} & 3.0668   & 0.0004 & 0.0098 & 0.5060 &0.4823 & 0.1481 &  0.0067 & 0.0005 & 1.2150\\
                  1_{3} & 4.8057   & 0.0022 & 0.5060 & 1.4629 &2.6995 & 0.0218 &  0.3418 & 0.0659 & 5.1001\\
                  1_{5} & 5.1080^* & 0.0022 & 0.4823 & 2.6995 &1.6023 & 0.2780 &  0.1540 & 0.0105 & 5.2288\\
                  1_{4} & 5.4724   & 0.0003 & 0.1481 & 0.0218 &0.2780 &11.5766 &  3.0925 & 1.7817 & 16.8990\\
                  1_{6} & 5.6332^* & 0.0001 & 0.0067 & 0.3418 &0.1540 & 3.0925 &  1.2682 & 0.4129 & 5.2762\\
                  1_{7} & 7.0280^* & 0      & 0.0005 & 0.0659 &0.0105 & 1.7817 &  0.4129 & 0.8250 & 3.0965\\
\hline  &
 \text{sum} & 0.1508 & 1.2150 & 5.1001 &  5.2288 &16.8990 & 5.2762 & 3.0965 & 36.9051
\end{array}$\\

\raggedright
\newpage
\begin{figure}[!htbp]
\centering
{\includegraphics[width = 6in]{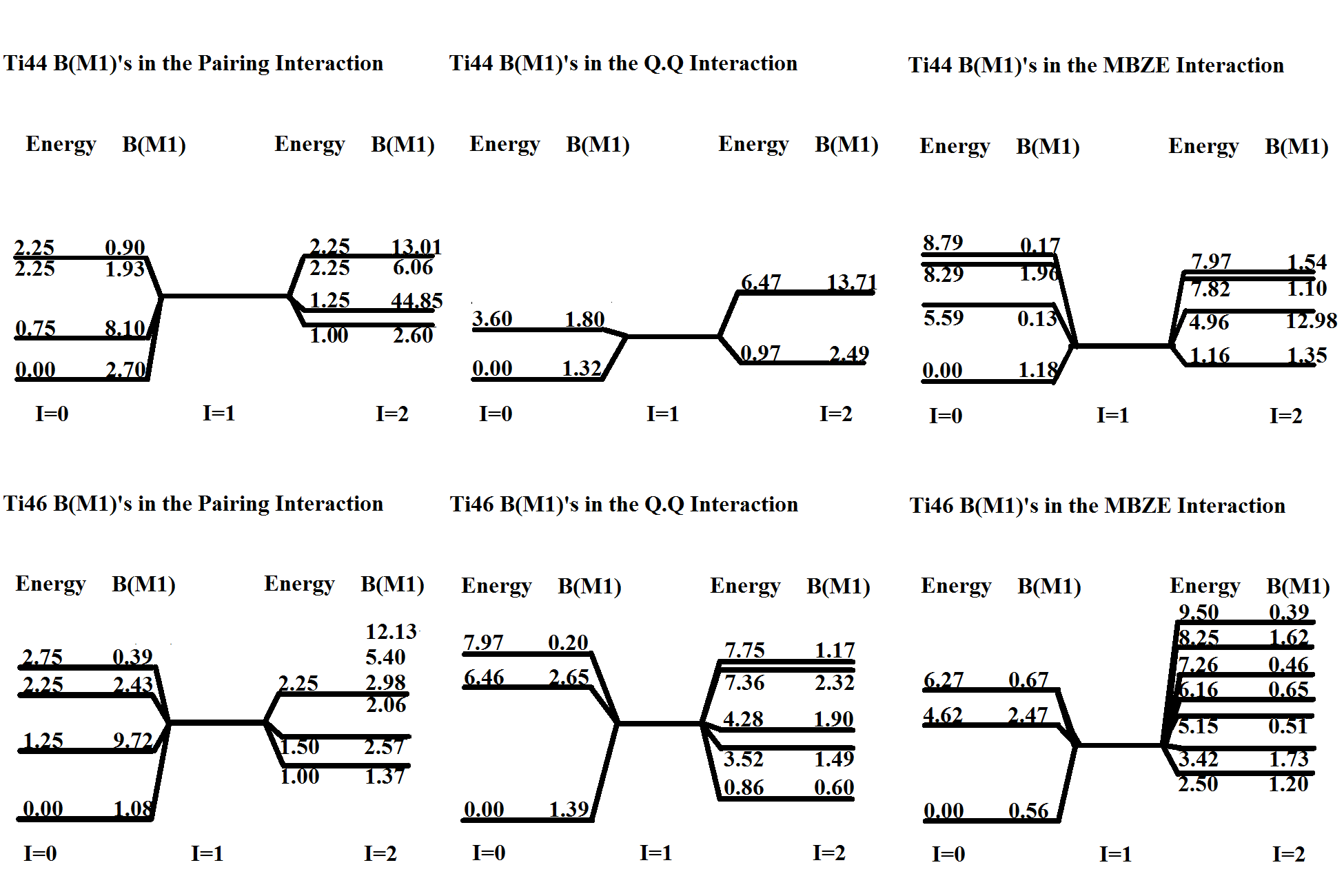}}
\caption{Strong B(M1) Diagrams}
\end{figure}

We here repeat the expressions for the B(M1)'s given by Harper and Zamick [12]

\begin{gather}
B(M1)=\frac{3}{4\pi}\frac{2I_{f}+1}{2I_{i}+1}\left[g_{j_{p}}X_{1}+(-1)^{I_{f}-I_{i}}g_{j_{n}}X_{2}\right]^{2}\\
\text{Here, }g_{j}=g_{l}\pm\left\lbrace \frac{g_{s}-g_{l}}{2l+1}\right\rbrace \\
\begin{align}g_{s_{p}}=5.586 &  & g_{l_{p}}=1\\
g_{s_{n}}=-3.826 &  & g_{s\l_{n}}=0
\end{align}
\end{gather}

For the case $T_{f}$ is not equal to $T_{i}$ we find: 
\begin{gather}
X_{1}=(-1)^{I_{f}-I_{i}+1}X_{2}\\
B(M1)=\frac{3}{4\pi}\frac{2I_{f}+1}{2I_{i}+1}(g_{jp}-g_{jn})^{2}X_{1}^{2}
\end{gather}

\section{Selection Rules for the Pairing Interaction}

In a previous work we commented on selection rules for vanishing B(M1)'s
with a J=0 T=1 pairing interaction. The basis states were written
($v,T,t$)-seniority, isospin and reduced isospin. We briefly repeat
the selection rules here and refer to tables II,III,IV, and V. For
the $J=0$ $T=1$ pairing interaction we previously found the following:

a. Transitions with $\Delta T$=2 or more are forbidden.

b. For N=Z nuclei $T$=1 to $T$=1 M1 transitions are zero.

c. $\Delta v$=4 M1 transitions are forbidden.

d. Transitions in which both $v$ and $t$ change are forbidden.

Here we discuss case d in more detail than we did in ref 9. In say,
$^{46}$Ti we break the six nucleons into three pairs. We cannot have
an M1 transition involving only a pair of identical particles--we
must consider a neutron-proton pair. The only way to change seniority
is to create or destroy a $J$=0 $T$=1 pair. The reduced isospin excludes
$J$=0 $T$=1 pairs. If the M1 operator acts on a $J$=0 $T$=1 pair it creates
a $J$=1 $T$=0 pair. Since this new pair has $T$=0 it will not change the
reduced isospin. Alternately, if we act on a $J$=1 $T$=0 pair we note that
because it has $T$=0, it does not contribute to the reduced isospin.
The M1 operator will change this to a $J$=0 $T$=1 pair and such pairs
are excluded from the reduced isospin set. Hence, if we change seniority
we cannot change the reduced isospin. These arguments of course also
explain c, why $v$ cannot change by more than two units.

In tables IV and V, we show using the $J$=0 pairing interaction, $^{46}$Ti
transitions from 1$^{+}$ to 0$^{+}$ states and 1$^{+}$ to 2$^{+}$
states respectively. We find an abundance of confirmations of rule
d. We note in table IV that all transitions from the $J$= 0$^{+}$
(220) configuration to 1$^{+}$ states except for (220) vanish. These
latter 1$^{+}$ states have configurations (411), (611) and (421).
In table V we see that B(M1)'s from $J$=2$^{+}$ (410) to 1$^{+}$
(611) vanish.

In table V we also see that $\Delta v$=4 B(M1)'s are zero e.g. (211)
to (611). Note that the $\Delta T$=2 transitions from the last 2$^{+}$
state (231) to $J$=1$^{+}$ $T$=1 states all vanish. This selection
rule is the easiest to understand i.e. in terms of the Wigner-Eckart
theorem.

\section{Selection Rules for the Q.Q and $J_{max}$ T=0 Interactions.}

We find also some vanishing B(M1)'s when the quadrupole-quadrupole
interaction Q.Q is employed. Some are not surprising like the $T$=1
to $T$=1 transitions in N=Z $^{44}$Ti shown in Table VII. Likewise,
the $\Delta T$=2 transitions in Table IX from the 2$_{18}$ $T$=3
state in $^{46}$Ti to all $T$=1 $J$=1$^{+}$ states. However, the
vanishing B(M1)'s in the top line of Table VIII involving $J$=0$^{+}$
and $J$=1$^{+}$ states in $^{46}$Ti are hard to explain and we
will not attempt to do so here. The vanishings are from the lowest
0$^{+}$ state to two $T$=1 states and one $T$=2 state. But we have
non-vanishings to other $T$=1 and $T$=2 states, so there is no simple
connection with isospin. 

There are no vanishings for the latter states except of course in
the bottom row where we have the $\Delta T$=2 selection rule . That
is to say the 0$_{6}$ state has $T$=3 and will not connect to $J$=1$^{+}$
$T$=1 states.

There are other peculiarities with the Q.Q interaction. As noted in
\cite{Harper}, in $^{44}$Ti there is a degenerate pair of $J$=2$^{+}$states
at 7.75 MeV--one has isospin $T$=0 and the other $T$=2. 

Likewise we find some hard to understand selection rules for the $J_{max}$
$T$=0 interaction. In Table XVI we find for $^{46}$ Ti, from the
lowest 0$^{+}$ state there are vanishing B(M1)'s to one $T$=1 state
and two $T$=2 states. As in the case with Q.Q this is hard to understand.

\section{Sums of Sums}

Note that the sum of sums, i.e. sum of all B(M1)'s from all 1$^{+}$states
to all 0$^{+}$ states, is independent of the interaction--same for
pairing as for Q.Q

This is easy to show, utilizing the fact that the $D$'s form a
complete set and the wave functions are normalized to unity. 
\begin{gather}
\sum_{\alpha}D^{\alpha}(J_{p}J_{n})D^{\alpha}(J_{p}^{\prime}J_{n}^{\prime})=\delta_{J_{p}J_{p}^{\prime}}\delta_{J_{n}J_{n}^{\prime}}\\
\sum_{J_{p}J_{n}}D^{\alpha}(J_{p}J_{n})^{2}=1
\end{gather}

This leads to the following expression for the sum of sums. 
\begin{gather}
SS=\frac{3}{4\pi}\frac{2I_{f}+1}{2I_{i}+1}(g_{p}-g_{n})^{2}\sum_{J_{p}J_{n}}U(1,J_{p}I_{f}J_{n};J_{p}I_{i})^{2}\times J_{p}(J_{p}+1)
\end{gather}

\section{Non-Monotonic Behavior of the B(M1) 1$_{1}$ to 0$_{1}$ as One Switches
from $J=0$ pairing to $J_{max}$ Pairing}

Let us focus on the 1$_{1}^{+}$ transitions. The conventional scissors
mode excitation is from 0$_{1}^{+}$ to 1$_{1}^{+}$ which will of
course be a factor of three larger than the reverse transition 1$_{1}^{+}$
to 0$_{1}^{+}$. With the Q.Q interaction we note however that there
are even larger B(M1)'s to other states. In $^{46}$Ti whereas the
B(M1) for 1$_{1}$ to 0$_{1}$ is 1.3901, from 1$_{1}$ to 0$_{2}$
it is 2.6505, almost twice as large. One possible explanation of this
is that the 0$_{2}$ state is a double scissors mode excitation.

Let us however now focus on the 1$_{1}^{+}$ to 0$_{1}^{+}$ in $^{46}$Ti,
i.e. the conventional spin-scissors mode. Here are some values from
the above tables:
\subsection*{Table XXIII. Comparison of 1$_{1}^{+}$ to 0$_{1}^{+}$ in $^{46}$Ti}
\centering $\begin{array}{ccc}
\hline \text{Interaction} & \text{Table} & \text{B(M1)}\\
\hline J=0\text{ Pairing } & \text{VI} & 1.0799\\
\text{ Q.Q } & \text{VII} & 1.3901\\
\text{ MBZE } & \text{XII} & 0.55962
\\\hline \end{array}$\\
 \raggedright

It is puzzling that Q.Q and MBZE are so different because there is
a big overlap between their respective wave functions.

To better understand this we now consider simple interactions which
are mixtures of $J$=0 pairing and $J_{max}$ pairing: 
\begin{gather}
V=a\delta_{J=0}+b\delta_{J=7}
\end{gather}

We present the B(M1) for selected values of $(a,b)$,
\subsection*{Table XXIV. B(M1) for a Mixture of Pairing and $J_{max}$}
\centering $\begin{array}{cccc}
\hline \text{a} & \text{b} & \text{B(M1)}\\
\hline -1 & 0 & 1.082 & \text{\ensuremath{J}=0 pairing}\\
-1.15 & -1 & 0.210 & \text{close to lowest B(M1)}\\
-1 & -1 & 0.260 & \text{equal \ensuremath{J}=0 and \ensuremath{J=J_{max}} pairing}\\
0 & -1 & 1.641 & \text{\ensuremath{J_{max}}pairing}
\\\hline \end{array}$\\
 \raggedright

We see a fairly complicated behavior--relatively large B(M1)'s at
the two limits, $J$=0 pairing and $J_{max}$ pairing. However, for
equal $J$=0 and $J=J_{max}$=7 pairing the value is much smaller
0.210 as compared with 1.080 and 1.641 We get a non-monotonic behavior
for this spin scissors mode. We get the lowest possible B(M1) for
$(a,b)$ close to (-1.15,-1) i.e. B(M1)=0.210.

Going back to Q.Q and MBZE, evidently there is more $J=0$ and $J=J_{max}$
interference in MBZE than there is in Q.Q.

\section{Additional Comments}

We note that the B(M1) from the lowest 1$^{+}$ to the lowest 0$^{+}$
(generally considered the scissors mode transition) is considerably
smaller than the transition from this 1$^{+}$ to all 0$^{+}$ states.
For example with MBZE (the most realistic interaction here) the respective
numbers are 1.6533 and 3.6136. The respective numbers from 1$^{+}$
to 2$^{+}$ are 1.0252 and 6.6738.

Note in Table XIV that along the diagonal of the one to one ``transitions''
in $^{44}$Ti the values of ``B(M1)'' are all the same. Of course
they are not real transitions, but they can be related to the magnetic
moments. Note that for N=Z nuclei in the single $j$ approximation
the magnetic $g$ factor is independent of the details of the wave
function. As seen in the appendix of \cite{Harper} the value is 
\begin{gather}
g=\frac{g_{p}+g_{n}}{2}=0.55
\end{gather}
This explains why all the diagonal ``B(M1)'s'' are the same in $^{44}$Ti.
This is not the case in $^{46}$Ti. The off diagonal zeros in Table
XV are due to the fact, as mentioned in \cite{Harper} that in N=Z nuclei
transitions from $T$ to the same $T$ (in this case T=1) are forbidden. 

\end{document}